\documentclass[reprint, superscriptaddress, aps, prl]{revtex4-2}

\usepackage{graphicx}
\usepackage{dcolumn}
\usepackage{bm}
\usepackage{hyperref}

\usepackage[utf8]{inputenc}
\usepackage[american,british]{babel}
\usepackage[T1]{fontenc}
\usepackage{bm}
\usepackage{amsmath,amsthm,amssymb} 
\usepackage{bbold}     
\usepackage{color}
\usepackage{soul} 
\usepackage{orcidlink}
\usepackage{braket}

\newcommand{\im}{{\rm i}}
\newcommand{\smt}{Science, Mathematics and Technology Cluster, Singapore University of Technology and Design, 8 Somapah Road, 487372 Singapore}
\newcommand{\epd}{Engineering Product Development Pillar, Singapore University of Technology and Design, 8 Somapah Road, 487372 Singapore}
\newcommand{\cqt}{Centre for Quantum Technologies, National University of Singapore 117543, Singapore}
\newcommand{\astar}{Quantum Innovation Centre, Agency for Science, Technology and Research, Singapore 138634, Singapore}
\newcommand{\mitrle}{Research Laboratory of Electronics, Massachusetts Institute of Technology, Cambridge, Massachusetts 02139, USA} 
\newcommand{\mitnse}{Department of Nuclear Science and Engineering, Massachusetts Institute of Technology, Cambridge, Massachusetts 02139, USA}
\newcommand{\mitphy}{Department of Physics, Massachusetts Institute of Technology, Cambridge, Massachusetts 02139, USA}
\newcommand{\ictp}{The Abdus Salam International Center for Theoretical Physics (ICTP), I-34151 Trieste, Italy}
\newcommand{\napoli}{Dipartimento di Fisica, Universit\`a di Napoli ``Federico II'', I-80126 Napoli, Italy}
\newcommand{\catania}{Dipartimento di Fisica e Astronomia “Ettore Majorana”, Universit\`{a} di Catania, Via S. Sofia 64, 95123 Catania, Italy}
\newcommand{\infn}{INFN, Sezione di Catania, I-95123 Catania, Italy}

\begin{document}

\title{Entanglement transition in unitary system-bath dynamics} 

\author{Bo Xing~\orcidlink{0000-0002-9651-5505}}
\email{boxing92@mit.edu}
\affiliation{\mitrle}
\affiliation{\astar}
\affiliation{\smt}
\author{Giuliano Chiriac\`{o}~\orcidlink{0000-0002-3906-4437}} 
\affiliation{\catania}
\affiliation{\infn}
\author{Paola Cappellaro~\orcidlink{0000-0003-3207-594X}}
\affiliation{\mitrle}
\affiliation{\mitnse}
\affiliation{\mitphy} 
\author{Rosario Fazio~\orcidlink{0000-0002-7793-179X}} 
\affiliation{\ictp} 
\affiliation{\napoli}
\author{Dario Poletti~\orcidlink{0000-0001-5626-0176}}
\email{dario_poletti@sutd.edu.sg}
\affiliation{\smt}
\affiliation{\epd}
\affiliation{\cqt} 
\date{\today}

\begin{abstract} 
The evolution of a system coupled to baths is commonly described by a master equation that, in the long-time limit, yields a steady-state density matrix.
However, when the same evolution is unraveled into quantum trajectories, it is possible to observe a transition in the scaling of entanglement within the system as the system–bath coupling increases — a phenomenon that is invisible in the trajectory-averaged reduced density matrix of the system.
Here, we go beyond the paradigm of trajectories from master equations and explore whether a qualitatively analogous entanglement-scaling transition emerges in a single unitary evolution of the combined system–bath setup, without monitoring the dynamics of the system.
We investigate the scaling of entanglement in a unitary quantum setup composed of a $2$D lattice of free fermions, where each site is coupled to a fermionic bath. 
As the system–bath coupling increases, the logarithmic fermionic negativity reveals an entanglement transition from logarithmic-law to area-law scaling.
This occurs while the system's steady-state properties are trivial, highlighting that the signatures of these different scalings are within the bath-bath correlations. 
Evidence of the transition is also found in the mutual information and the correlations of the full system–bath setup, suggesting that the entanglement transition is underpinned by a change in the spatial structure of quantum information.
\end{abstract}

\maketitle

\textit{Introduction ---}
Open quantum systems are central to quantum information science and underpin the development of quantum technologies. In a myriad of experimentally relevant scenarios, such systems are accurately modeled by the Gorini–Kossakowski–Sudarshan–Lindblad (GKSL) equation ~\cite{Lindblad1976, GoriniSudarshan1976} (see~\cite{Manzano2020, StefaniniMarino2025} for a pedagogical introduction). In a many-body setting, furthermore, new collective phenomena appear due to the interplay between the unitary evolution of the system and its interaction with the external environment. Widely studied examples of this sort are, just to mention a few of them, dissipative phase transitions, many-body state preparation, exotic states of matter in driven-dissipative systems, and dissipatively engineered topological properties~\cite{Daley2014, FazioSchiro2025}.
\begin{figure}[htbp!]
    \includegraphics[width=\linewidth]{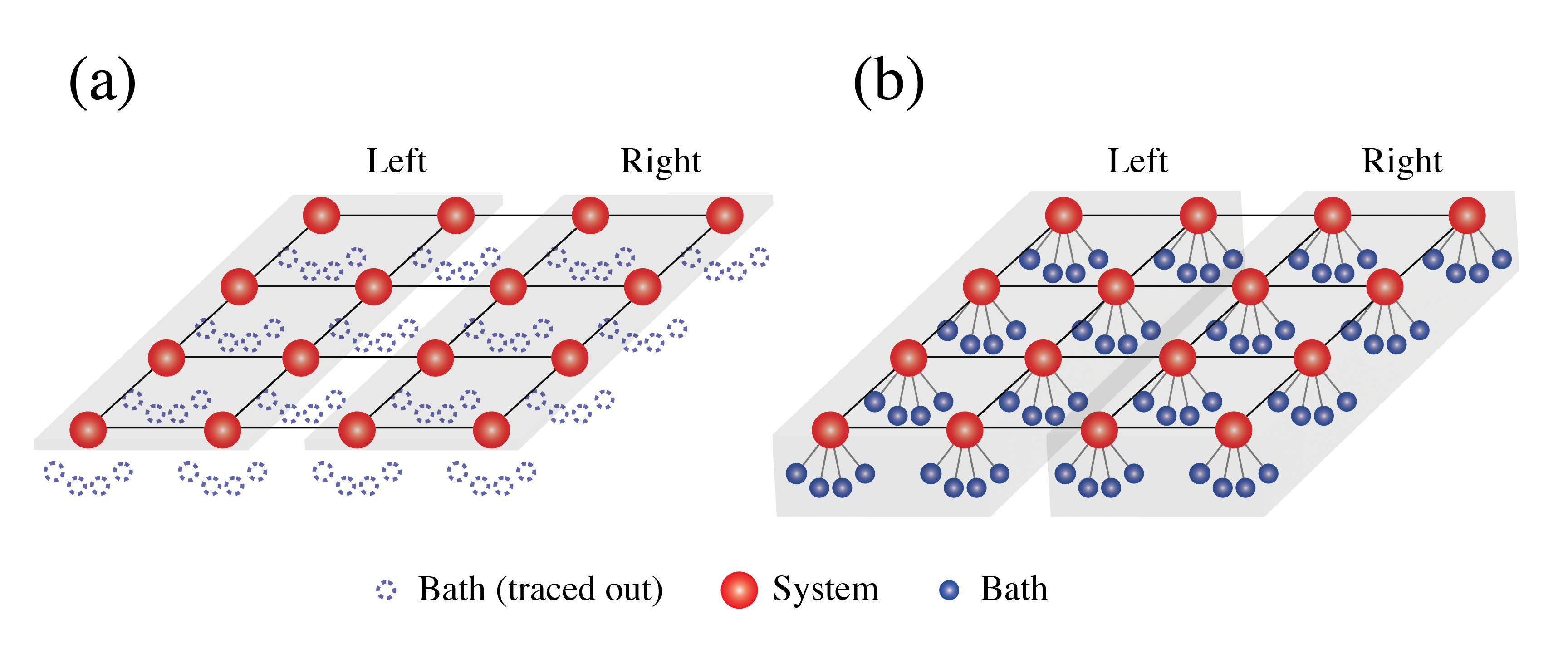}
    \caption{
    (a) A $2$D lattice of free fermions with open boundary conditions is coupled to external baths.
    After tracing out the baths, the dissipative dynamics of the system can be approximated by the GKSL master equation.
    For further analysis, we partition the system into the left and right partitions, each consisting of $N/2$ system sites.
    (b) We can also consider a unitary evolution, in which the environment is modeled as $M$ discrete local bath modes coupled to each system site.
    The left and right partitions each contain $N/2 \times (M+1)$ system and bath sites, and the degrees of freedom of the baths are retained.
    }
    \label{fig:M_Schematics}
\end{figure}

All the cases mentioned above can be described by considering the dynamics of the reduced density matrix of the system. 
If, however, one is able to keep a record of the interaction events of the system with the baths, e.g, monitor the dynamics along single trajectories~\cite{DalibardMolmer1992}, the picture becomes even richer. 
There are phenomena that can be revealed in monitored dynamics, while washed out when considering the reduced density matrix.
A prominent example is the entanglement dynamics of monitored many-body systems, which have recently attracted significant interest and can lead to a transition called the measurement-induced phase transition (MIPT)~\cite{LiFisher2018, SkinnerNahum2019, SzyniszewskiSchomerus2019, LiFisher2019, SzyniszewskiSchomerus2020, PotterVasseur2022, LuntPal2022, FisherVijay2023}.
This stems from the interplay between the unitary dynamics that often spread entanglement across the system and the monitoring that can suppress the spreading.

MIPTs are not only of theoretical interest for quantum statistical mechanics and information theory, but also have practical implications for quantum error correction, quantum computation, and the design of quantum devices, where controlling entanglement is essential~\cite{ChoiAltman2020, FidkowskiHastings2021}.
In $1$D systems, the MIPT is well-established and has been extensively studied in both free and interacting systems~\cite{CaoDeLuca2019, BaoAltman2020, GullansHuse2020, FujiAshida2020, GullansHuse2020a, AlbertonDiehl2021, BaoAltman2021, TurkeshiSchiro2021, Sierant2022dissipativefloquet, CoppolaCollura2022, GalSchiro2023, GranetDreyer2023, FavaNahum2023, LoioTurkeshi2023, XingPoletti2024, PoboikoMirlin2025}, where the bipartite entanglement exhibits different scaling behaviors depending on the measurement probability or measurement strength.
This has also been verified experimentally at small scales~\cite{NoelHuse2022, Google2023, KohMinnich2023}.
For larger systems, experimental studies are limited by the post-selection problem, whereby one needs to select trajectories that correspond to specific sequences of measurement outcomes. 
Significant progress has also been made beyond 1D systems~\cite{TurkeshiDalmonte2020, LuntPal2021, JianLudwig2022, LiuChen2022, SierantTurkeshi2022}, with numerical evidence of MIPT observed even in free fermion systems~\cite{TurkeshiDalmonte2020, LuntPal2021, JianLudwig2022, LiuChen2022, SierantTurkeshi2022, PoboikoMirlin2024}.
Studies have also shown MIPTs in non-Markovian setups~\cite{ChiriacoDalmonte2023, TsitsishviliChiriaco2024, MuzziChriaco2025}, and how one can generalize MIPTs as an information exchange symmetry breaking~\cite{KellyMarino2025, KellyMarino2025b}. 

In the above studies, MIPTs have been understood mostly through frameworks that rely on an ensemble of quantum trajectories~\cite{GardinerZoller2004, WisemanMilburn2009, Jacobs2014, Breuer2022}.
However, quantum trajectories are not the only framework capable of retaining information about system–bath interactions.
A more complete approach is to study the joint unitary evolution of the combined system and baths, where the bath degrees of freedom are retained and preserve the information that would otherwise be lost when probing only the system.
The system could still reach a steady trivial state, but the interaction between the system and baths can leave traces of the overall evolution in the baths, which could potentially be used to extract information.
We thus ask: can an entanglement transition also manifest in this unitary system–bath setup, in the absence of monitoring?
This would provide important insights into the dynamics of many-body open quantum systems and offer a way to probe entanglement transitions without suffering from the post-selection problem.

We address this question by studying a $2$D lattice of free fermions coupled to external baths.
We first study this dissipative system using a quantum-trajectory unraveling of the GKSL master equation, which reveals an MIPT at finite measurement rate.
We then consider the fully unitary evolution of the combined system–bath setup, without trajectories or monitoring.
In this unitary case, we observe an entanglement transition as the system–bath coupling strength is varied, as evidenced by the logarithmic fermionic negativity.
This is possible because the unitary setup retains the bath degrees of freedom, which preserve information about the dynamics.
Notably, the transition is also visible in the mutual information and in the correlations of the baths, which require significantly fewer resources to evaluate.

\textit{Entanglement Transition in Trajectories ---}
We consider a $2$D lattice of free fermions with $N = L \times L$ system sites and open boundary conditions.
The system Hamiltonian is
\begin{equation}\label{eq:H_S}
    H_{\text{S}} = - \sum_{\langle n,n'\rangle} J ( a^\dagger_n a^{}_{n'} + h.c. ) + \sum_nh_{\text{s}} a^\dagger_n a^{}_n, 
\end{equation}
where $a^{\dagger}_{n}$($a^{}_{n}$) is the fermionic creation (annihilation) operator of the system at site $n$, $J$ is the tunneling magnitude between nearest sites $\langle n,n'\rangle$ in the system, and $h_{\text{s}}$ is the on-site energy of the system. 

The system is coupled to external baths, which induce dissipative dynamics.
In many physically relevant situations, one can accurately model this setup by tracing out the bath degrees of freedom and deriving an effective reduced description of the system alone, as depicted in Fig.~\ref{fig:M_Schematics}(a). 
Under the Born–Markov and secular approximations, the resulting dynamics of the reduced system density matrix $\rho$ is governed by the GKSL master equation~\cite{Lindblad1976,GoriniSudarshan1976},
\begin{equation}\label{eq:GKSL}  
    % \frac{d\rho}{dt} = -\im [H_{\text{S}}, \rho] + \sum_n \frac{\gamma}{2}\left( \mathcal{D}_{a^\dagger_n}(\rho) + \mathcal{D}_{a_n}(\rho) \right)
    \frac{d\rho}{dt} = -\im [H_{\text{S}}, \rho] + \gamma\sum_k \left( L_{k}^{} \rho L_{k}^\dagger - \frac{1}{2} \left\{ L_{k}^\dagger L_k^{}, \rho \right\} \right)
,\end{equation}
where $\rho$ is the reduced density operator of the system, $\gamma$ is the relaxation rate, $L_k$ are the jump operators given by $a^{}_{n}$ and $a^{\dagger}_{n}$, and the sum over $k$ ensures that both operators are used on each site of the system. We are working in units such that $\hbar = 1$.

Given that Eqs.~(\ref{eq:H_S}, \ref{eq:GKSL}) are quadratic, the dynamics of the system can be described in terms of the correlation matrix $C^{\text{ME}}=\langle a^{\dagger}_{n} a^{}_{n^\prime} \rangle$.
Following Eq.~(\ref{eq:GKSL}), the reduced system correlation matrix $C^{\text{ME}}$ converges to a featureless infinite-temperature steady state at sufficiently long time $t_{\text{s}} = 50/J$, i.e., $C^{\text{ME}}(t_{\text{s}}) \approx \mathbb{1}/2$.
However, the trajectory unraveling of the GKSL master equation reveals an MIPT in the bipartite entanglement between the left and right partitions shown in Fig.~\ref{fig:M_Schematics}(a).
To demonstrate this, we use a quantum-jump unraveling of the GKSL master equation and simulate $N_{\text{TJ}} = 256$ stroboscopic trajectories. 
Starting from an initial checkerboard configuration, each trajectory undergoes unitary evolution under $H_{S}$ for a time step $dt = 0.1$, followed by a quantum jump $a_n \;(a_n^\dagger)$ occurring with probability $p_{\pm} = \gamma dt$~\cite{supp}.  
To quantify the bipartite entanglement, we compute the von Neumann entropy $S(x) = - \sum_{\lambda} \left[ \lambda \log \lambda + (1 - \lambda) \log (1 - \lambda) \right]$, where $\lambda$ are the eigenvalues of the correlation matrix $x$.
We also calculate the logarithmic fermionic negativity~\cite{supp}, $E(x)$, which is an entanglement monotone for both pure and mixed states~\cite{AudenaertEisert2003, Plenio2005}

\begin{figure}[htbp!]
    \includegraphics[width=\linewidth]{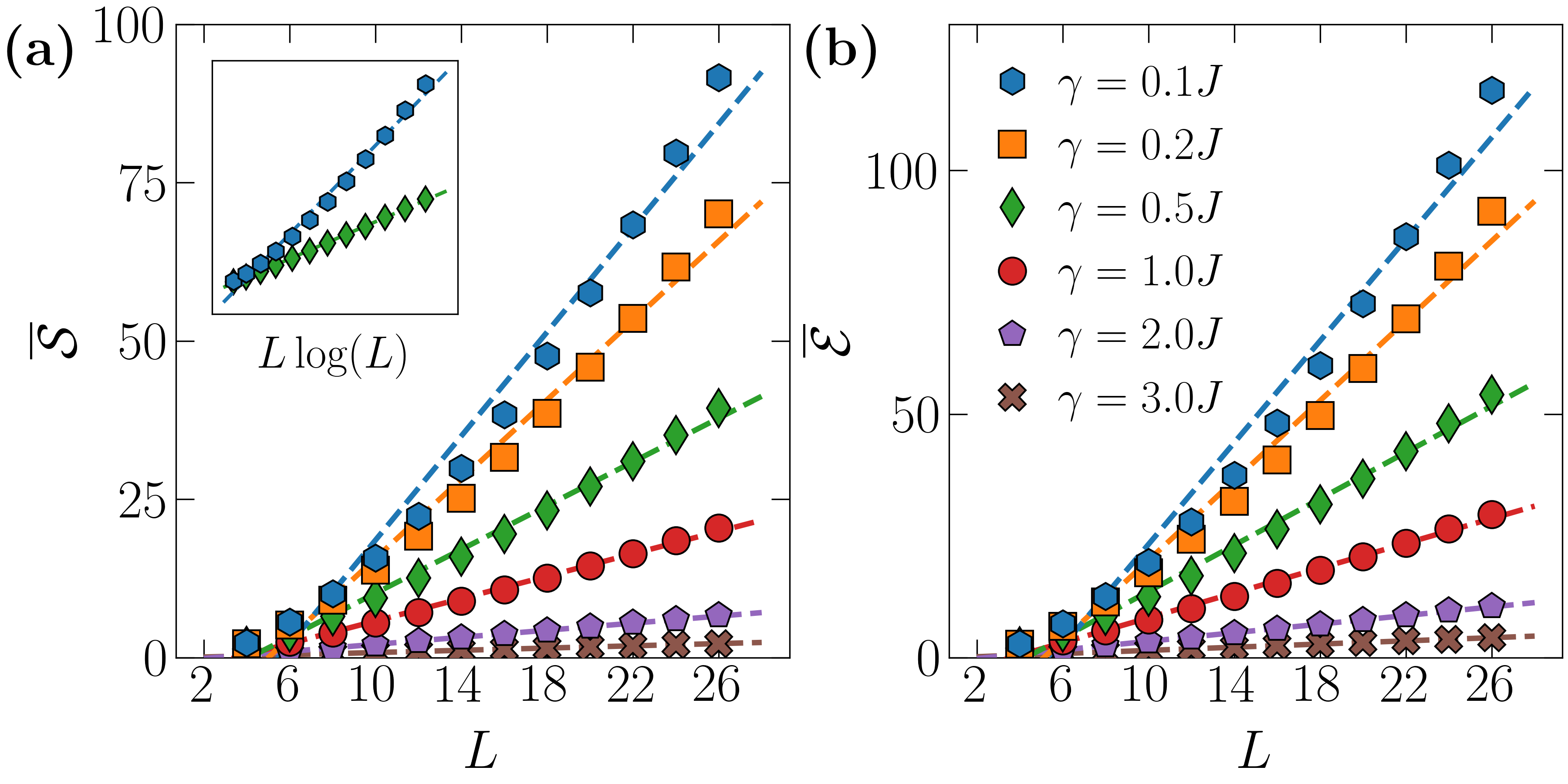}
    \caption{The trajectory-averaged steady-state (a) bipartite von Neumann entropy $\overline{\mathcal{S}}$ and (b) bipartite logarithmic fermionic negativity $\overline{\mathcal{E}}$ as a function of $L$.
    In the inset of (a), we plot $\overline{\mathcal{S}}$ against $L\log{L}$ for $\gamma=0.1J$ and $0.5J$.
    The filled symbols are numerical results, and the dashed lines are the linear best-fit lines based on all data points.
    The error bars are smaller than the symbol sizes and thus not shown.}
    \label{fig:M_TJ}
\end{figure}

In Fig.~\ref{fig:M_TJ}(a), we plot the trajectory-averaged steady-state bipartite von Neumann entropy, $\overline{\mathcal{S}}=  \sum_\alpha S\left(\mathrm{Tr}_R C^\alpha(t_{\text{s}})\right) / N_{\text{TJ}}$, where $C^\alpha$ is the correlation matrix of the system along trajectory $\alpha$ and $\mathrm{Tr}_R$ is the partial trace over the right partition.
In Fig.~\ref{fig:M_TJ}(b) we show the trajectory-averaged steady-state bipartite logarithmic fermionic negativity, $\overline{\mathcal{E}} = \sum_\alpha E\left(C^\alpha(t_{\text{s}})\right) / N_{\text{TJ}}$.
In the main panels, both $\overline{\mathcal{S}}$ and $\overline{\mathcal{E}}$ scale linearly with the partition boundary size $L$ when $\gamma \geq 2.0J$, a clear signature of area-law scaling in 2D.
As $\gamma$ decreases, the scaling becomes superlinear, indicating a transition away from area-law scaling.
In the inset of Fig.~\ref{fig:M_TJ}(a), we find that $\overline{\mathcal{S}}$ at $\gamma=0.1J$ and $0.5J$ scales linearly with $L\log{(L)}$ instead, a signature of logarithmic-law scaling in $2$D.

\textit{Entanglement Transition in Unitary Dynamics ---}
We consider a unitary model in which each system site is coupled to $M$ bath modes, as shown in Fig.~\ref{fig:M_Schematics}(b).
In the appropriate limits, tracing out the baths recovers the GKSL dynamics of the system considered above.
The total system-bath Hamiltonian is
\begin{align}
    H_{\text{tot}} &= H_{\text{S}} + H_{\text{B}} + H_{\text{SB}},  \label{eq:H_tot} \\
    H_{\text{B}}          &= \sum_n^N \sum_m^M \omega^{}_{m} b^\dagger_{n,m}b^{}_{n,m}, \label{eq:H_B} \\
    H_{\text{SB}}         &= - \sum_n^N \sum_m^M V(\omega^{}_m) ( a^\dagger_{n} b^{}_{n,m} + H.c ), \label{eq:H_SB}
\end{align}
where $b^{\dagger}_{n,m}$($b^{}_{n,m}$) is the fermionic creation (annihilation) operator of the $m-$th bath mode at site $n$, $\omega_{m} = m\omega_{\text{max}}/M$ is the energy of the $m-$th bath mode (independent of $n$), and $V(\omega_{m})$ is the magnitude of the system-bath coupling.

Since $H_{\text{tot}}$ is quadratic in the fermionic operators, the unitary evolution preserves Gaussianity.
The total correlation matrix
\begin{equation}
    C_0 =
    \begin{bmatrix}
        C^{\text{SS}} & C^{\text{SB}} \\
        {C^{\text{SB}}}^\dagger & C^{\text{BB}} \\
    \end{bmatrix}
,\end{equation}
evolves unitarily in time $t$ under $U(t) = e^{- i H_{\text{tot}} t}$. 
Here, $C^{\text{SS}}_{n,n'} = \langle a_n^\dagger a_{n'} \rangle$ is the system correlation matrix,
$C^{\text{BB}}_{(n,m),(n',m')} = \langle b_{n,m}^\dagger b_{n',m'} \rangle$ is the bath correlation matrix,
and $C^{\text{SB}}_{n,(n',m)} = \langle a_n^\dagger b_{n',m} \rangle$ encodes system–bath correlations.
To recover $C^{\text{SS}} = C^{\text{ME}}$, $M$ and $\omega_{\text{max}}$ must be sufficiently large, and the system-bath coupling has to be $V(\omega_{m}) = \sqrt{\gamma \frac{\omega_{m}}{M} \frac{\omega_{\text{max}}}{\pi h_{\text{s}}}}$~\cite{deVegaWolf2015}.
For all unitary system-bath evolutions, we use $\omega_{\text{max}}=10J$, $h_s=5J$, and $M=100$~\cite{supp}.

At $t = 0$, $C^{\text{SS}}$ is in a checkerboard configuration and $C^{\text{BB}}$ is the infinite-temperature mixed state.
Since the total system-bath evolution is unitary and therefore deterministic, the steady-state correlation matrix can be obtained from a single evolution.
At $t \gtrsim t_{\text{s}}$, $C^{\text{SS}}$ converges to the same featureless infinite-temperature state as $C^{\text{ME}}$, but the bath contains non-trivial correlations that are built up through the interaction with the system~\cite{supp}.

To study the entanglement between the left and right partitions of the unitary setup, we reorder the elements of the correlation matrix $C_0$ as  
\begin{equation}
    C = 
    \begin{bmatrix}
        C^{\text{LL}} & C^{\text{LR}} \\
        {C^{\text{LR}}}^\dagger & C^{\text{RR}} \\
    \end{bmatrix} ,
\end{equation}
where $C^{\text{LL}}$ ($C^{\text{RR}}$) is the correlation matrix of the left (right) partition, and $C^{\text{LR}}$ is the correlation matrix between the left and right partitions. 
We compute the unitary-evolved steady-state bipartite logarithmic fermionic negativity, $\mathcal{E} = E(C(t_{\text{s}}))$ and introduce two additional complementary steady-state quantities: the mutual information~\cite{CalabreseCardy2004, AmicoVedral2008}, $\mathcal{I} = S\left( C^{\text{LL}}(t_{\text{s}}) \right) + S\left( C^{\text{RR}}(t_{\text{s}}) \right) - S \left( C(t_{\text{s}}) \right)$, and the connected correlation weight betwen the left and right partitions, $\mathcal{C} = \sum_{(n,m),(n^\prime,m^\prime)} \left| C^{\text{LR}}_{(n,m),(n^\prime,m^\prime)}(t_{\text{s}}) \right|^2$.
The mutual information provides an upper bound to the connected correlation weight~\cite{WolfIgnacio2008, LeporiTrombettoni2022}.
Both quantities exhibit similar phenomenology but are numerically less expensive than logarithmic fermionic negativity, allowing one to probe larger system sizes.

\begin{figure}[htbp!]
    \includegraphics[width=\linewidth]{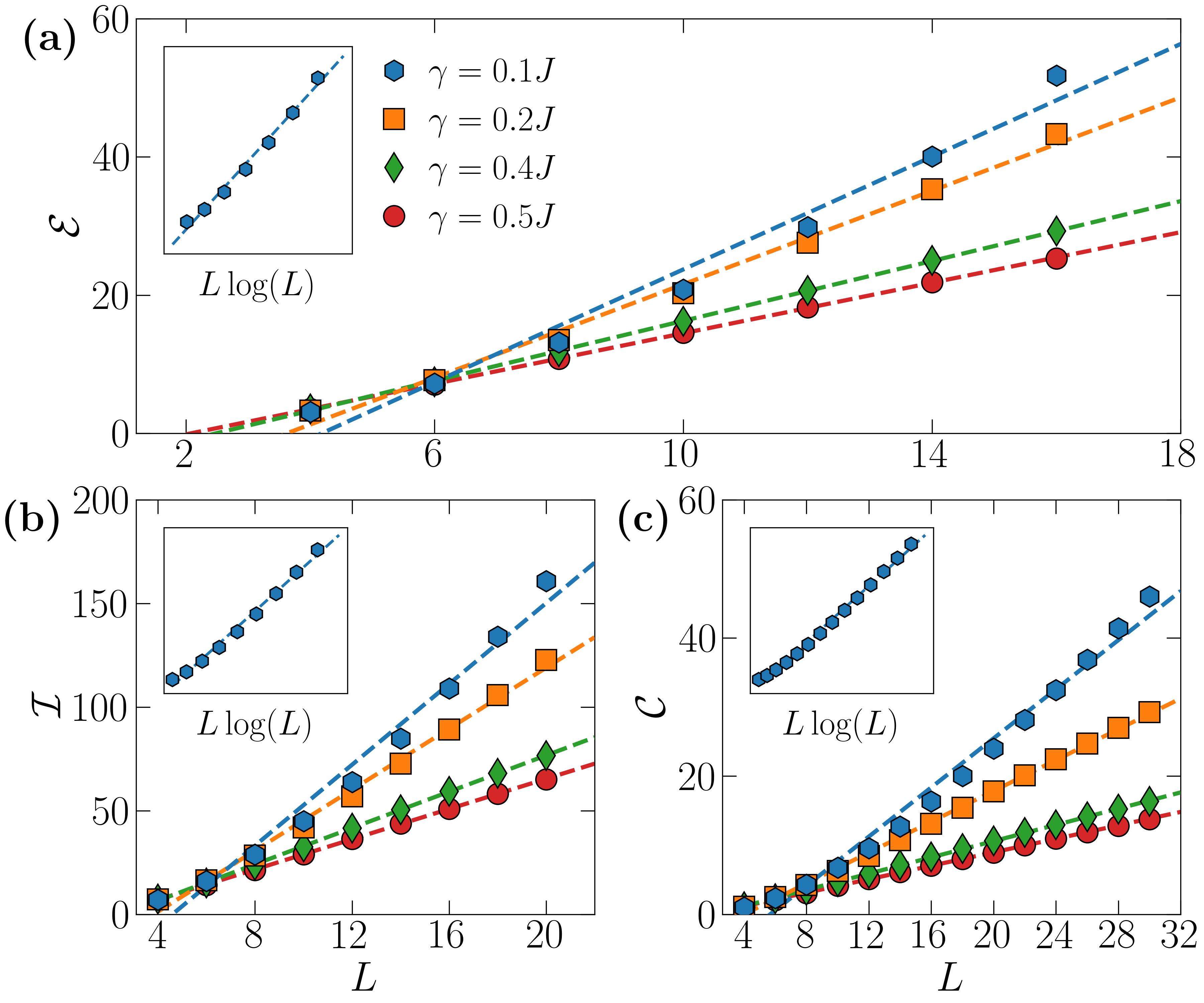}
    \caption{The unitary-evolved steady-state (a) bipartite logarithmic fermionic negativity $\mathcal{E}$, (b) mutual information $\mathcal{I}$, and (c) the connected correlation weight between the left and the right partitions $\mathcal{C} $ as a function of $L$.
    In the insets, we plot the same quantities against $L\log{L}$ for $\gamma=0.1J$.
    The filled symbols are numerical results, and the dashed lines are linear best-fit lines based on all data points. 
    }
    \label{fig:M_U}
\end{figure}

In Fig.~\ref{fig:M_U}, we plot the unitary-evolved steady-state (a) bipartite logarithmic fermionic negativity between the left and right partitions $\mathcal{E}$, (b) mutual information between the left and right partitions $\mathcal{I}$, and (c) connected correlation weight between the left and right partitions $\mathcal{C}$.
In the main panels, the results for $\gamma \geq 0.4J$ exhibit a clear area-law scaling, where all three quantities scale linearly with $L$.
At $\gamma = 0.1J$, the scaling is distinctly superlinear with $L$.
In the insets, we investigate the scaling behavior at $\gamma = 0.1J$ and find that it follows a logarithmic-law scaling, increasing linearly with $L\log(L)$.

\begin{figure}[htbp!]
    \includegraphics[width=\linewidth]{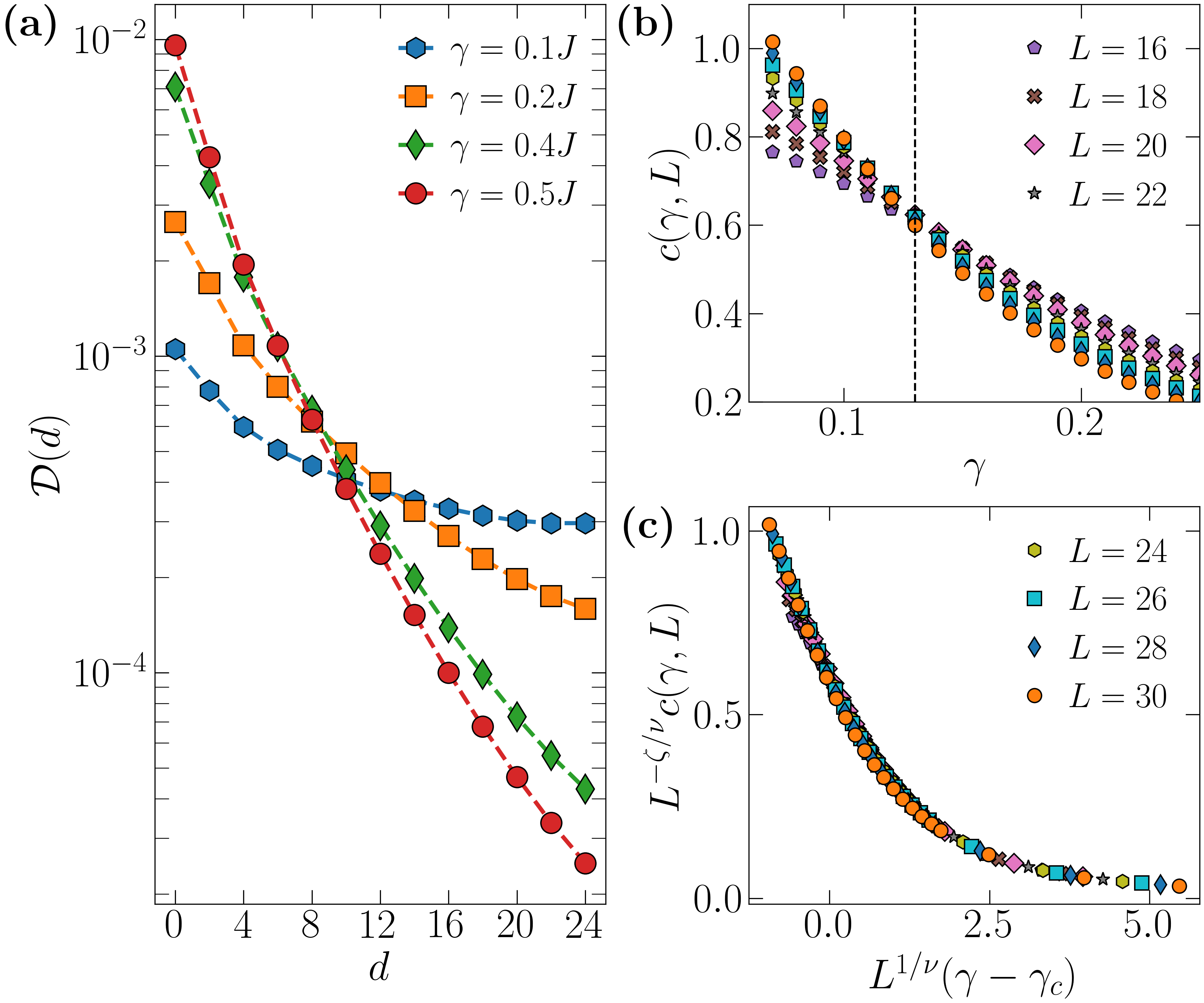}
    \caption{(a) The steady-state density-density correlation as a function of the distance $\mathcal{D}(d)$ for different $\gamma$ values.
    The system size is $N=676$.
    (b) $c(\gamma, L)$ obtained by fitting $\mathcal{C}(\gamma, x) = c(\gamma, L)x \ln{(x)} + b(\gamma, L)x$ within the range $x \in [8, L]$.
    The vertical dashed line marks the estimated size-invariant phase transition point $\gamma_c$.
    (c) Finite size scaling analysis of $c(\gamma, L)$.
    The data show a good collapse with exponent $\nu = 1.26 \pm 0.48$, $\zeta = 0.00 \pm 0.01$, and $\gamma_c = (0.13 \pm 0.01)J$.
    Subplots (b, c) share the same legends.
    }
    \label{fig:M_Corr}
\end{figure}

To further understand the physics of the entanglement transition in the unitary system-bath setup, we analyze the steady-state spatial structure of bath–bath correlations, which contribute the most towards the connected correlation weight~\cite{supp}.
Specifically, we define the steady-state density-density correlation between baths at sites $n$ and $n^\prime$ as 
\begin{align} \label{eq:corr}
    \mathcal{D}_{n, n^\prime} &= \sum_{m, m^\prime} \big|\langle D_{n,m} D_{n^\prime, m^\prime} \rangle - \langle D_{n,m} \rangle \langle D_{n^\prime, m^\prime} \rangle \big| \nonumber \\
    &= \sum_{m, m^\prime} \big |C^{\text{BB}}_{(n, m), (n^\prime, m^\prime)} \big|^2
,\end{align}
where $D_{n,m} = b_{n,m}^\dagger b_{n,m}$ is the density of the bath~\footnote{For $n = n^\prime$, we neglect contributions from $m = m^\prime$ to exclude trivial self-correlations of the bath modes. The simplification of Eq.~(\ref{eq:corr}) is done using Wick's theorem.}.
The density-density correlation as a function of the distance $d$ is therefore
\begin{equation}
    \mathcal{D}(d) = \frac{1}{N_{\text{pairs}}(d)} \sum_{\mathrm{dist}(n,n') = d} \mathcal{D}_{n,n'}
,\end{equation}
where $N_{\text{pairs}} (d)$ is the number of $(n, n^\prime)$ pairs sharing the same city block distance $\text{dist}(n, n^\prime) = d$.
In Fig.~\ref{fig:M_Corr}(a), we plot $\mathcal{D}(d)$ for different $\gamma$ values.
We find that the logarithmic-law to area-law entanglement transition is underpinned by a change in the spatial structure of bath–bath correlations.
At smaller values of $\gamma = 0.1J$, baths at different sites can establish long-range correlations through their respective system sites, whereas increasing $\gamma$ suppresses these correlations to short distances.

Lastly, we show that the different scaling regimes are separated by a genuine phase transition and determine the corresponding critical point $\gamma_c$.
For each $\gamma$, we obtain $b(\gamma, L)$ and $c(\gamma, L)$ by fitting the connected correlation weight $\mathcal{C}(\gamma, x) = c(\gamma, L) x \ln (x) + b(\gamma, L) x$ within different ranges of linear system sizes, $x \in [8, L]$.
We use $\mathcal{C}$ for the fitting because it can be evaluated for larger system sizes (up to $L = 30$ in this work), enabling more accurate finite-size scaling. 
As shown in Fig.~\ref{fig:M_U}, $\mathcal{C}$ behaves qualitatively similar to $\mathcal{I}$ and $\mathcal{E}$.
In the presence of a phase transition, finite-size scaling theory~\cite{FisherBarber1972, CampostriniVicari2014} predicts that both $c(\gamma, L)$ and $b(\gamma, L)$ become size-independent at the critical point. 
This property has been used to determine the critical point in monitored free-fermion systems~\cite{ChahineBuchhold2024, PoboikoMirlin2024, MuzziChriaco2025}.

In Fig.~\ref{fig:M_Corr}(b), we show that $c(\gamma, L)$ exhibits clear size invariance around $\gamma \sim 0.14J$.
For $\gamma < 0.14J$, the logarithmic correction coefficient $c(\gamma, L)$ increases as the system size grows, whereas for $\gamma > 0.14J$ it decreases with increasing system size, revealing two distinct scaling regimes.
We perform a finite size scaling (FSS) analysis with the pyfssa package~\cite{Melchert2009, Sorge2015} and fit the behavior of $c(\gamma, L)$ with the function $c(\gamma, L) = L^{\zeta/\nu} f\left(L^{1/\nu} (\gamma - \gamma_c)\right)$, where $\zeta,\ \nu$ are the critical exponents, and $\gamma_c$ is the critical point in the thermodynamic limit.
In Fig.~\ref{fig:M_Corr}(c), we observe that the curves for different system sizes collapse nicely onto each other, yielding critical exponents $\zeta = 0.00 \pm 0.01$, $\nu = 1.26 \pm 0.48$, and $\gamma_c = (0.13 \pm 0.01)J$.
Obtaining $\zeta \approx 0.00$ means that $c(\gamma, L)$ does not scale with $L$ in the thermodynamic limit.
The value of $c(\gamma, \infty)$ is finite for $\gamma < \gamma_c$ and becomes $0$ for $\gamma > \gamma_c$.
A similar FSS analysis on the area-law coefficient $b(\gamma, L)$ gives $\zeta = 0.00 \pm 0.01$, $\nu = 0.99 \pm 0.52$, and $\gamma_c = (0.13 \pm 0.02)J$, within the error bars of the estimates obtained from $c(\gamma, L)$.
Together, these results provide strong evidence of an entanglement phase transition in our unitary system-bath setup.

\textit{Conclusion ---}
In this work, we demonstrate that the entanglement phase transition -- usually observed from unraveling a GKSL master equation -- is strikingly revealed also in a fully unitary evolution of a system–bath setup, without additional external monitoring or trajectories. 
For the unitary system-bath setup evolution, we show that the steady-state bipartite logarithmic fermionic negativity $\mathcal{E}$, mutual information $\mathcal{I}$, and connected correlation weight between the left and right partitions $\mathcal{C}$, all exhibit a transition from logarithmic-law to area-law scaling as the system-bath coupling rate $\gamma$ increases. 
As the steady state of the system is a trivial infinite-temperature state, the observed transition can only be witnessed by studying information about the dynamical evolution stored in the baths.
These results provide a new perspective on entanglement transitions in open quantum systems. 

The numerical value of $\gamma_c$ at which the entanglement scaling transitions to area-law differs significantly between the unitary evolution of the system-bath setup and the trajectory unraveling of the GKSL master equation. 
One possible reason is that the GKSL equation integrates out bath degrees of freedom, thereby discarding bath correlations that play a central role in the entanglement transition.
Furthermore, trajectory-level properties have additional dependence on the choice of unraveling scheme~\cite{PiccittoRossini2024}.

Beyond the theoretical insight, this work also has interesting experimental implications.
The unitary system-bath setup provides new opportunities to study entanglement transitions using quantum simulators.
Unlike the trajectory approach, which requires post-selection of measurement outcomes, the unitary setup starts from an initial mixed-state and yields the steady-state total correlation matrix from a single deterministic evolution.
The same matrix can also be reconstructed from random pure bath initializations, see~\cite{supp}.

One interesting future direction is to extend the investigation to interacting systems, where volume law entanglement can be fully developed.
In the presence of interactions, the system dynamics is no longer Gaussian, and the numerical complexity increases exponentially, thus potentially requiring a quantum simulator implementation.
In this case, accurate simulation of interacting baths can be achieved with a relatively small number of qubits, as shown both theoretically and experimentally~\cite{XuPoletti2022, XuPoletti2023, ZhangPoletti2024}, and one could consider transition witnesses such as coherence or correlations, which can be evaluated more efficiently than entanglement. 
Work along this direction is currently underway.

\textit{Acknowledgments ---}
We thank B.K. Agarwalla, V. Khemani, M. Kulkarni, A. Stasiuk, and R.R.W. Wang for the helpful discussions.
B.X. was supported by the A*STAR International Fellowship. 
We acknowledge support from the Singapore Ministry of Education grant MOE-T2EP50123-0017, and from the Centre for Quantum Technologies grant CQT$\_$SUTD$\_$2025$\_$01.
This work was in part supported by the National Science Foundation under Grants No. PHY1915218, No. PHY1734011.
G.C. acknowledges support from the Italian Ministry of University and Research under Project PNRR-MUR No. E63C22001000006-ICSC.
R.\,F. acknowledges support from  ERC under grant agreement n.101053159 (RAVE), and by the PNRR MUR project PE0000023-NQSTI. 
Views and opinions expressed are those of the authors only and do not necessarily reflect those of the European Union or the European Research Council. 
Neither the European  Union nor the granting authority can be held responsible for them.
The computational work for this article was partially performed at the National Supercomputing Centre, Singapore \cite{nscc}.
The data that support the findings of this article are publicly available~\cite{data}.

\bibliographystyle{apsrev4-2}
\bibliography{main}

\clearpage
% \onecolumngrid
\begin{center}
\textbf{\large Supplemental Material}
\end{center}

\setcounter{section}{0}
\setcounter{equation}{0}
\setcounter{figure}{0}
\setcounter{table}{0}

\renewcommand{\thesection}{S\arabic{section}}
\renewcommand{\theequation}{S\arabic{equation}}
\renewcommand{\thefigure}{S\arabic{figure}}
\renewcommand{\thetable}{S\arabic{table}}

\section{Unraveling protocol}\label{App:unraveling}

Our system with Hamiltonian 
\begin{equation}
    H_{\text{S}} = - \sum_{\langle n,n'\rangle} J ( a^\dagger_n a^{}_{n'} + h.c. ) + \sum_nh_{\text{s}} a^\dagger_n a^{}_n, 
\end{equation}
undergoes a dissipative dynamics where particles are created and destroyed with a rate $\gamma$:
\begin{equation}\label{Eq:Lindblad}
    \dot\rho=-i[H_{\text{S}},\rho]+\gamma\sum_n(\mathcal{D}[a_n]\rho+\mathcal{D}[a_n^\dagger]\rho)
\end{equation}
where $\mathcal{D}[L]\rho=L\rho L^\dagger-\frac{1}{2}\{L^\dagger L,\rho\}$.

We want to unravel this dynamics in terms of quantum trajectories. To that end, we consider a stroboscopic dynamics, where the system evolves unitarily with $H_{\text{S}}$ for a time $dt=0.1$, after which the system is subject to an instantaneous quantum jump. The quantum jumps are described by the action of a set of Kraus operators $M_\mu$.

\subsection{Unitary dynamics}

During the unitary dynamics, the pure state of one quantum trajectory evolves as $\ket{\psi}\rightarrow e^{-iH\tau_u}\ket{\psi}$. The correlation matrix  $C_{ij}=\langle a_i^{\dagger}a_j\rangle$ evolves as $C_{ij}\rightarrow\bra{\psi}e^{iH\delta t}a_i^{\dagger}a_je^{-iH\delta t}\ket{\psi}=\sum_{i'j'}R_{ii'}^{\dagger}C_{i'j'}R_{j'j}$ where $R_{mn}$ is the matrix that describes the evolution operator. It can be found by diagonalizing the Hamiltonian in Fourier space, writing the evolution operator in such base and then transforming back to real space
\begin{equation}
    R_{mn}=\frac1L\sum_{\mathbf k}e^{2\pi i(\mathbf{ m}-\mathbf{n})\cdot \mathbf{k}/L-i2J[\cos(2\pi k_x/L)+\cos(2\pi k_y/L)]\delta t}
\end{equation}

In the limit $L\rightarrow\infty$, this can be written in terms of the Bessel functions~\cite{CoppolaCollura2022}, but for finite $L$ we can easily compute it numerically at a small computational cost.

\subsection{Jump dynamics}

At the end of each cycle of the unitary evolution, the system is subject to the quantum jumps part of the dynamics. The Kraus operators satisfy the normalization condition $\sum_\mu M^{\dagger}_\mu M_\mu=\mathbb1$ and lead to an update of the system wavefunction $\ket{\psi}\rightarrow\ket{\psi'}=\frac{M_\mu\ket{\psi}}{||M_\mu\ket{\psi}||}$ with probability given by the Born rule $||M_\mu\ket{\psi}||^2=\bra{\psi}M^{\dagger}_\mu M_\mu\ket{\psi}$.

In our case the unraveling proceeds separately with distinct sets of Kraus operators associated to the loss or gain of particles: $\{M^+_{0,n}=\sqrt{1-p}+(1-\sqrt{1-p})\hat a_n^{\dagger}a_n, M^+_{1,n}=\sqrt{p}a_n^{\dagger}\}$ for the creation jump operators and $\{M^-_{0,n}=1-(1-\sqrt{1-p})\hat a_n^\dagger a_n, M^-_{1,n}=\sqrt{p}a_n\}$ for the destruction jump operators, where $p=\gamma\delta t$.

The jump protocol is the following.
For the jump associated to $a_n^{\dagger}$:
\begin{itemize}
    \item With probability $p\langle a_na_n^{\dagger}\rangle$ the state is updated as $\ket{\psi}\rightarrow\ket{\psi}=\frac{a_n^{\dagger}\ket{\psi}}{\sqrt{\langle a_na_n^{\dagger}\rangle}}$.
    \item With probability $1-\delta p$ the state is updated as $\ket{\psi}\rightarrow\ket{\psi}=\frac{M^+_{0,n}\ket{\psi}}{\sqrt{1-\delta p}}$ where $\delta p=p\langle a_na_n^{\dagger}\rangle$.
\end{itemize}

In terms of the correlation matrix: with probability $p(1-C_{nn})$ we apply $a_n^{\dagger}$ and find
\begin{equation}
    C_{ij} \rightarrow C_{ij}+\frac{(\delta_{in}-C_{in})(\delta_{nj}-C_{nj})}{1-C_{nn}}
\end{equation}

With probability $1-\delta p=1-p(1-C_{nn})$ we apply $M^+_{0,n}$ and find
\begin{align}
C_{ij}\rightarrow C_{ij} &- \frac{p}{1-\delta p}C_{in}C_{nj} \\
                  \notag &+ \frac{C_{nn}\delta_{in}\delta_{nj}(1-\sqrt{1-p})^2}{1-\delta p} \\
                  \notag &+ \frac{\sqrt{1-p}(1-\sqrt{1-p})(\delta_{in}C_{nj}+C_{in}\delta_{nj})}{1-\delta p}
\end{align}

For the jump associated with $a_n$:
\begin{itemize}
    \item With probability $p\langle a_n^{\dagger}a_n\rangle$ the state is updated as $\ket{\psi}\rightarrow\ket{\psi}=\frac{a_n\ket{\psi}}{\sqrt{\langle a_n^{\dagger}a_n\rangle}}$.
    \item With probability $1-\delta p$ the state is updated as $\ket{\psi}\rightarrow\ket{\psi}=\frac{M^-_{0,n}\ket{\psi}}{\sqrt{1-\delta p}}$ where $\delta p=p\langle a_n^{\dagger}a_n\rangle$.
\end{itemize}

In terms of the correlation matrix, with probability $pC_{nn}$ we apply $a_n$ and find
\begin{equation}
C_{ij} \rightarrow C_{ij}-\frac{C_{in}C_{nj}}{C_{nn}}
\end{equation}

With probability $1-\delta p=1-pC_{nn}$ we apply $M^-_{0,n}$ and find
\begin{align}
C_{ij}\rightarrow C_{ij} &+ \frac{p}{1-\delta p}C_{in}C_{nj} \\
                  \notag &+ \frac{C_{nn}\delta_{in}\delta_{nj}(1-\sqrt{1-p})^2}{1-\delta p} \\
                  \notag &- \frac{(1-\sqrt{1-p})(\delta_{in}C_{nj}+C_{in}\delta_{nj})}{1-\delta p}
\end{align}

\section{Steady state of the system}
\begin{figure}[htbp!]
    \includegraphics[width=\linewidth]{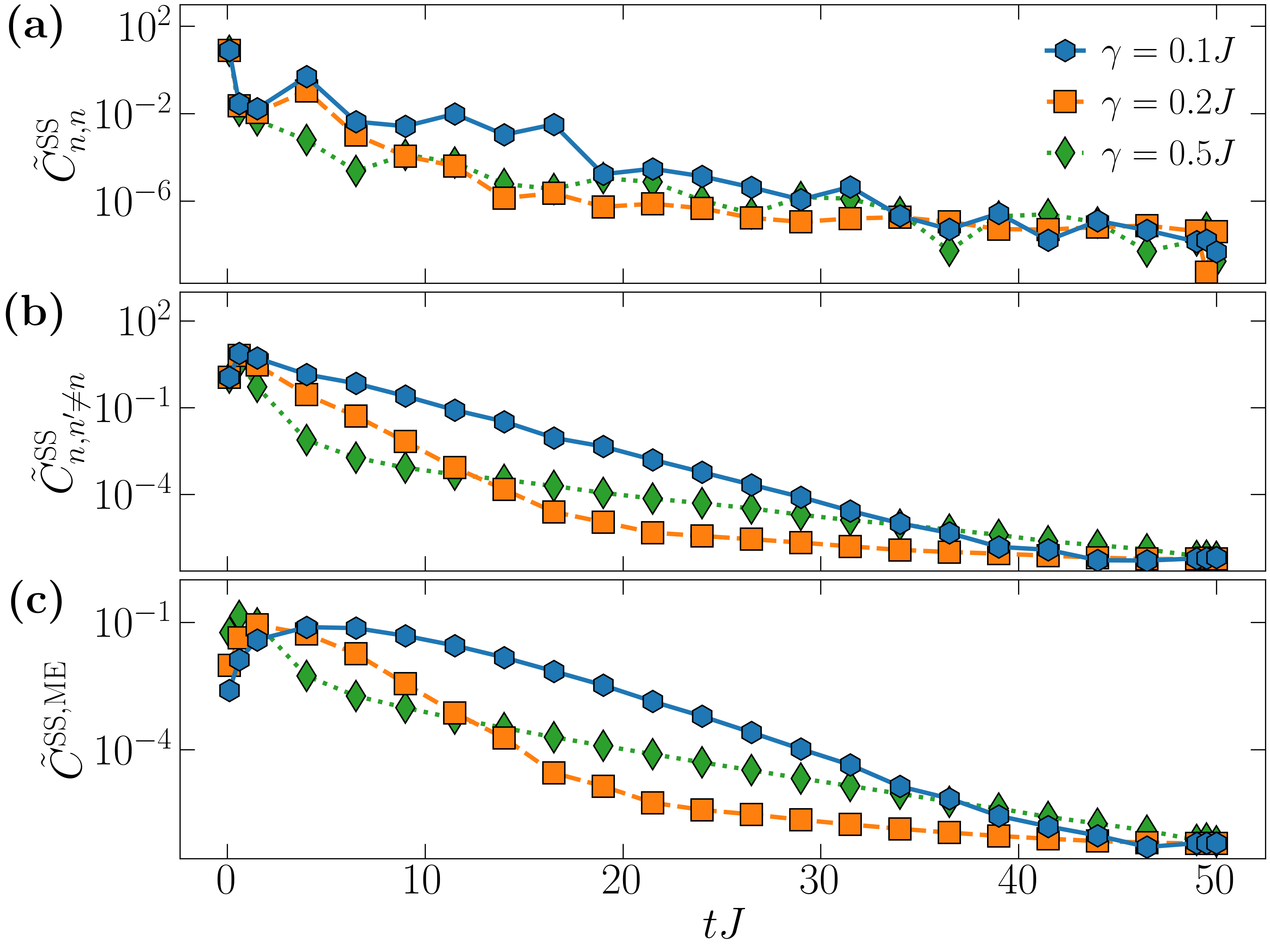}
    \caption{The dynamics of $C^{\text{SS}}$ in the unitary-evolved $C_0$. 
    (a) $\tilde{C}_{n,n}^{\text{SS}}$ quantifies the diagonal deviation from the infinite-temperature state. 
    (b) $\tilde{C}_{n,n^\prime}^{\text{SS}}$ quantifies the off-diagonal deviation from the infinite-temperature state. 
    (c) $\tilde{C}^{\text{SS}, \text{ME}}_{n, n^\prime}$ quantifies the difference between $C^{\text{SS}}$ and $C^{\text{ME}}$.
    In the steady-state, all three quantities are expected to diminish to $0$.
    The system size is $N=36$.}
    \label{fig:A_Dynamics}
\end{figure}

In the unitary system-bath setup, the total correlation matrix is
\begin{equation}
    C_0 =
    \begin{bmatrix}
        C^{\text{SS}} & C^{\text{SB}} \\
        {C^{\text{SB}}}^\dagger & C^{\text{BB}} \\
    \end{bmatrix}
,\end{equation}
where $C^{\text{SS}}$ is the system correlation matrix, $C^{\text{BB}}$ is the bath correlation matrix, and $C^{\text{SB}}$ is the system-bath correlation matrix.
On the other hand, the GKSL master equation allows access to only the system correlation matrix $C^{\text{ME}}$.
At large time $t_{\text{s}} = 50/J$, both $C^{\text{SS}}$ and $C^{\text{ME}}$ converge to the infinite-temperature state, i.e., $C^{\text{SS}}(t_{\text{s}}) = C^{\text{ME}}(t_{\text{s}}) = \mathbb{1}/2$, where $\mathbb{1}$ is the identity matrix.
We show that the unitary evolution of $C_0$ under the full system-bath Hamiltonian (Eq.(3) of the main text) yields the infinite-temperature steady state $C^{\text{SS}}$.
For all unitary system-bath evolutions, we use $\omega_{\text{max}}=10J$, $h_s=5J$, and $M=100$.
Increasing $M$ further does not quantitatively change the results.

At $t = 0$, $C^{\text{SS}}$ is initialized as a pure state with checkerboard occupation and $C^{\text{BB}}$ is initialized as an infinite-temperature mixed state.
We fixed the initial configuration of the system for ease of comparison between the two evolutions. 
The steady-state is independent of the initial system configuration.
In Fig.~\ref{fig:A_Dynamics}(a) we plot the diagonal deviation from the infinite-temperature state $\tilde{C}_{n,n}^{\text{SS}} = \sum_{n} | C^{\text{SS}}_{n,n} - 1/2|^2$, and in Fig.~\ref{fig:A_Dynamics}(b) we show the off-diagonal deviation $\tilde{C}_{n,n^\prime}^{\text{SS}} = \sum_{n,n^\prime} | C^{\text{SS}}_{n,n^\prime\neq n} |^2$.
In the steady state, these quantities are expected to vanish for both $C^{\text{SS}}$ and $C^{\text{ME}}$.
In Fig.~\ref{fig:A_Dynamics}(c), we quantify the difference between $C^{\text{SS}}$ and $C^{\text{ME}}$ using $\tilde{C}^{\text{SS}, \text{ME}} = \sum_{n,n^\prime} | C^{\text{ME}}_{n, n^\prime} - C^{\text{SS}}_{n, n^\prime}|^2$.
At short times, the difference is non-negligible due to the finite system-bath coupling effects neglected by the master equation. 
At longer time $t_{\text{s}} = 50/J$, the difference becomes $< 10^{-6}$ for all $\gamma$ values considered.   
Hence, under the unitary evolution of the system-bath setup, the system converges to the expected infinite-temperature state $C^{\text{SS}}=\mathbb{1}/2$.
Note that none of these quantities is averaged over the system size. Thus, the error per site is even smaller.

\section{Logarithmic Fermionic Negativity}\label{SM:LFN}
Given the total correlation matrix of the system-bath setup, $C$, we define $\Gamma = 2C - \mathbb{1}$.
For a bipartition of the total setup into left and right parts, $\Gamma$ is expressed as 
\begin{equation}
    \Gamma = 
    \begin{bmatrix}
        \Gamma^{LL} & \Gamma^{LR} \\
        {\Gamma^{LR}}^\dagger & \Gamma^{RR} \\
    \end{bmatrix} ,
\end{equation}
where each block corresponds to the correlations between the segments indicated in the superscript.
From $\Gamma$, we introduce the transformed matrices
\begin{equation}
    \Gamma_\pm = 
    \begin{bmatrix}
        \Gamma^{LL} & \pm i \Gamma^{LR} \\
        \pm i {\Gamma^{LR}}^\dagger & -\Gamma^{RR} \\
    \end{bmatrix} ,
\end{equation}
and
\begin{equation}
    \Gamma_* = \frac{1}{2} \left[ 1 - (1 + \Gamma_+ \Gamma_-)^{-1} (\Gamma_+ + \Gamma_-) \right] .
\end{equation}

Using these matrices, the logarithmic fermionic negativity is given by
\begin{equation} \label{eq:LFN}
    E(C) = \sum_j \left[ \ln \left( \sqrt{\mu_j} + \sqrt{1 - \mu_j} \right) + \frac{1}{2} \left( 1 - 2 \lambda_j + 2 \lambda_j^2 \right) \right],
\end{equation}
where $\mu_j$ and $\lambda_j$ are the eigenvalues of $\Gamma_*$ and $C$ respectively.

From Eq.~(\ref{eq:LFN}), we see that the calculation of the logarithmic fermionic negativity involves the diagonalization of the total correlation matrix $C$, regardless of the size of the bipartition we are interested in studying.
This is in contrast to the von Neumann entropy, which only requires the diagonalization of the correlation matrix of the partition of interest.

\section{Time Evolution of Various Observables}
\begin{figure}[htbp!]
    \includegraphics[width=\linewidth]{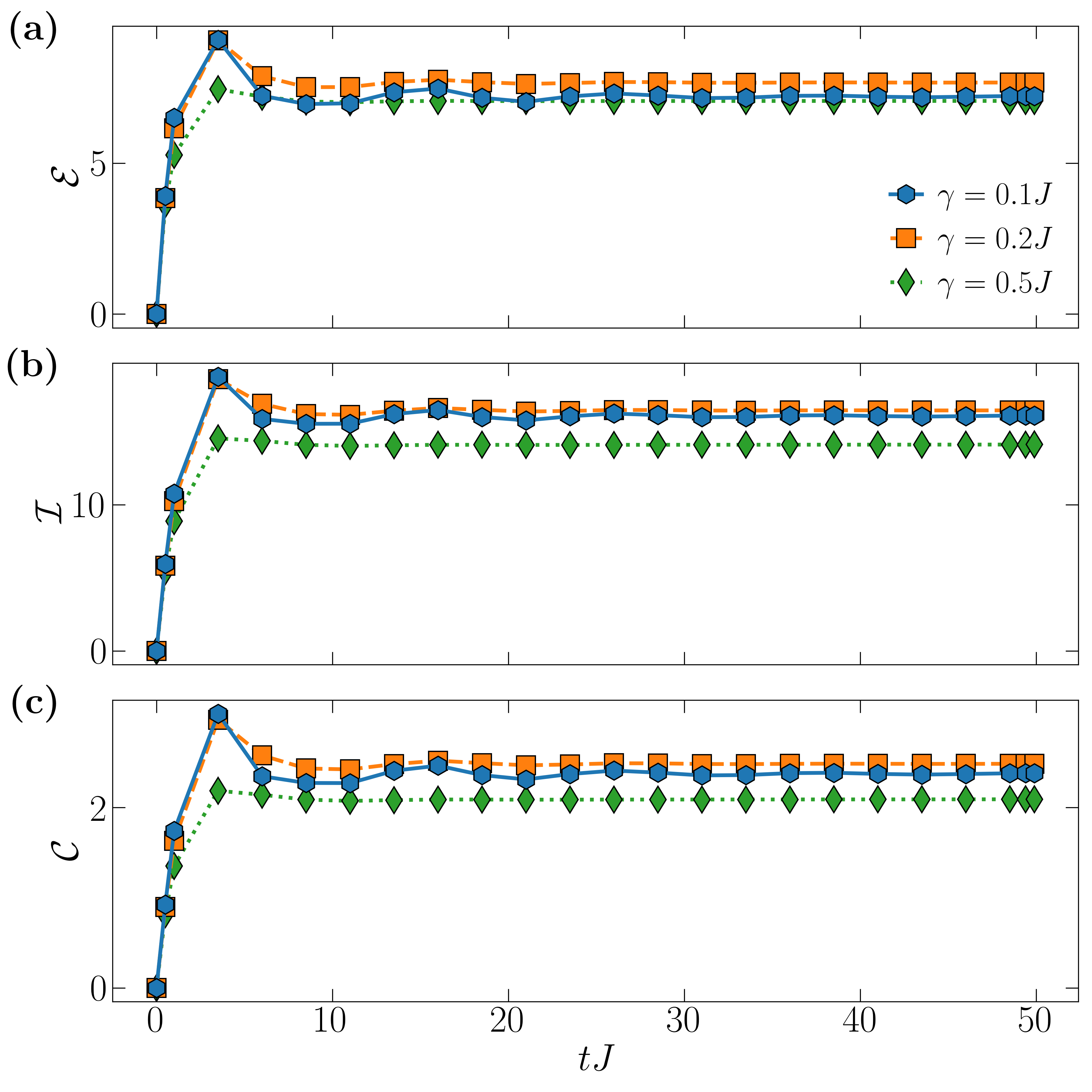}
    \caption{Time evolution of the entanglement related observables (a) $\mathcal{E}(t)$, (b) $\mathcal{I}(t)$, and (c) $\mathcal{C}(t)$.
    }
    \label{fig:A_Tevo}
\end{figure}

In Fig.~\ref{fig:A_Tevo}, we show the time evolution of the entanglement-related observables, (a) $\mathcal{E}(t)$, (b) $\mathcal{I}(t)$, and (c) $\mathcal{C}(t)$ in the unitary system-bath setup.
All three observables reach their steady-state values before $t_{\text{s}} = 50/J$.
The simulation parameters used are the same as in Fig.~\ref{fig:A_Dynamics}.

\begin{figure}[htbp!]
    \includegraphics[width=\linewidth]{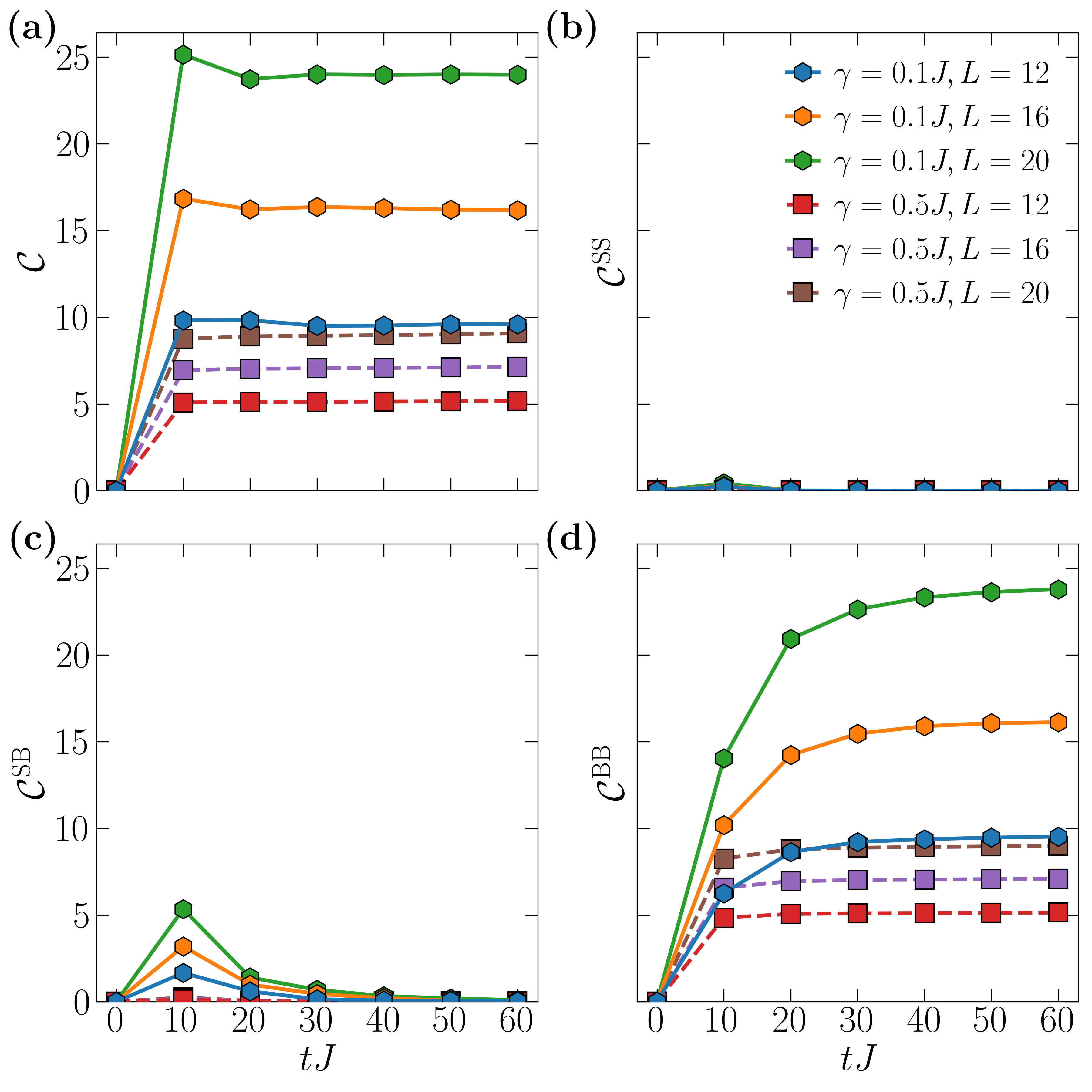}
    \caption{Time evolution of the connected correlation weight between (a) the left and right partitions, $\mathcal{C}$, (b) the left and right system, $\mathcal{C}^{\text{SS}}$, (c) between the left system and right baths, $\mathcal{C}^{\text{SB}}$, and (d) between the left and right baths, $\mathcal{C}^{\text{BB}}$.
    }
    \label{fig:A_off_diag}
\end{figure}

In Fig.~\ref{fig:A_off_diag}, we plot the connected correlation weight between (a) the left and right partitions, $\mathcal{C} = \sum_{(n,m),(n^\prime,m^\prime)} \left| C^{\text{LR}}_{(n,m),(n^\prime,m^\prime)} \right|^2$, (b) the left and right system, $\mathcal{C}^{\text{SS}} = \sum_{i \in L, i^\prime \in R} | C^{\text{SS}}_{i,i^\prime} |^2$, (c) between the left system and right baths, $\mathcal{C}^{\text{SB}} = \sum_{i \in L, i^\prime \in R, m} | C^{\text{SB}}_{i, (i^\prime, m)} |^2$, and (d) between the left and right baths, $\mathcal{C}^{\text{BB}} = \sum_{i \in L, m, i^\prime \in R, m^\prime} | C^{\text{BB}}_{(i,m), (i^\prime, m^\prime)} |^2$.
These quantities provide information on the entangling structure between different parts of the system-bath setup.
At $t \gtrsim t_{\text{s}}$, both $\mathcal{C}^{\text{SS}}$ and $\mathcal{C}^{\text{SB}}$ becomes negligible.
The correlations in $\mathcal{C}$ are mostly contributed by $\mathcal{C}^{\text{BB}}$, indicating that the entanglement between the left and right partitions originates entirely from bath-bath correlations.

\section{Unitary Evolution with Random Pure-State Baths}
\begin{figure}[htbp!]
    \includegraphics[width=\linewidth]{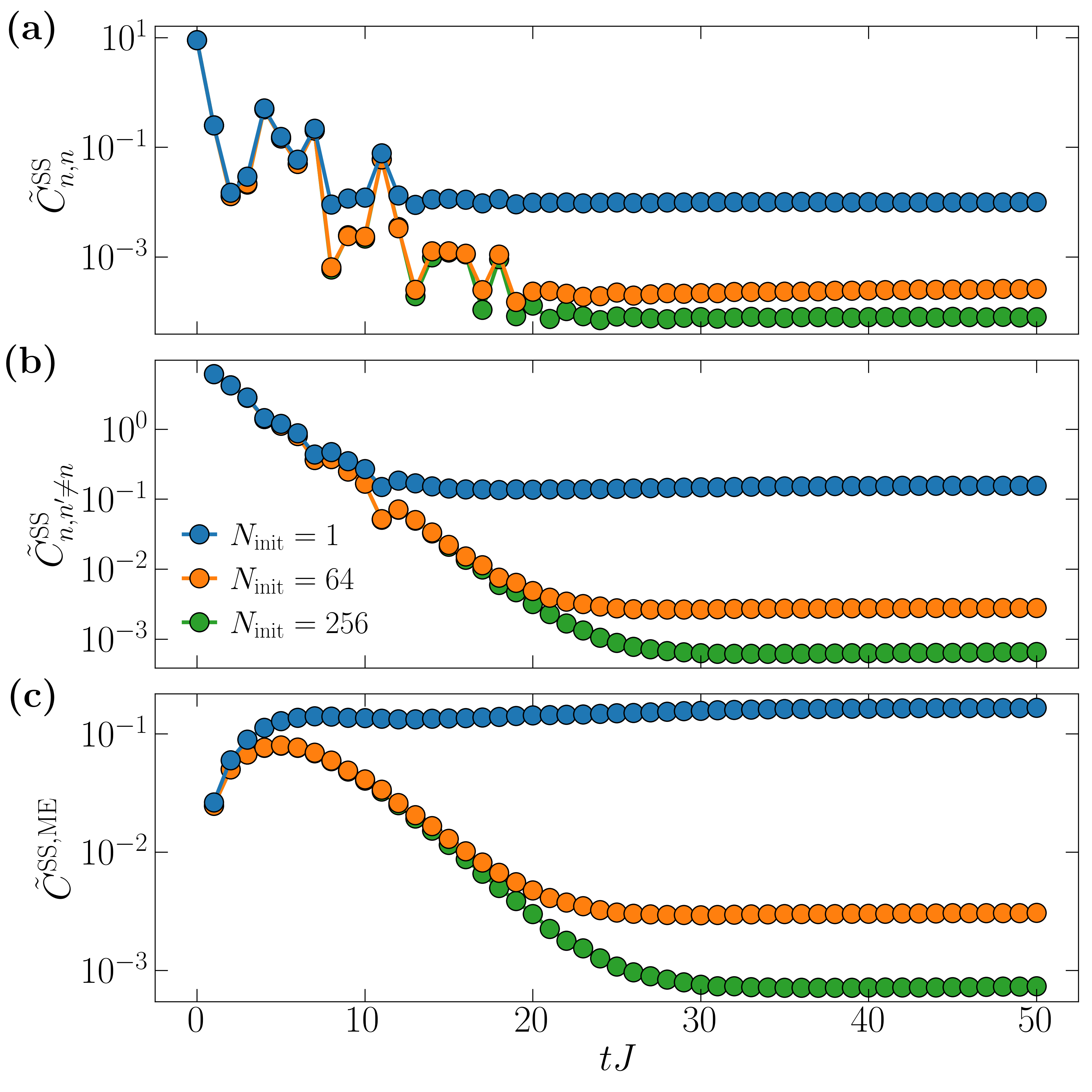}
    \caption{Time evolution of the system-system correlation matrix $\overline{C^{\text{SS}}}$ in $\overline{C_0}$. 
    (a) $\tilde{C}_{n,n}^{\text{SS}}$ quantifies the deviation of system densities from the infinite-temperature state. 
    (b) $\tilde{C}_{n,n^\prime}^{\text{SS}}$ quantifies the deviation of the system correlations from the infinite-temperature state. 
    (c) $\tilde{C}^{\text{SS}, \rm{ME}}_{n, n^\prime}$ quantifies the deviation of $\overline{C^{\text{SS}}}$ from $C^\text{ME}$ predicted by the GKSL master equation.
    The system size is $N=36$.}
    \label{fig:A_U_traj}
\end{figure}

In this section, we show that the correct system correlation matrix can also be recovered from our unitary setup with baths initialized randomly in pure states. 
Using the $H_{\text{tot}}$ (Eq.(3) of the main text), we evolve the total correlation matrix $C_0$ and compute the averaged $\overline{C_0}$ over $N_{\text{init}}$ different pure-state bath initializations.
In Fig.~\ref{fig:A_U_traj}(a) we plot the diagonal deviation from the infinite-temperature state $\tilde{C}_{n,n}^{\text{SS}} = \sum_{n} | \overline{C^{\text{SS}}_{n,n}} - 1/2|^2$.
In Fig.~\ref{fig:A_U_traj}(b) we show the off-diagonal deviation $\tilde{C}_{n,n^\prime}^{\text{SS}} = \sum_{n,n^\prime} | \overline{C^{\text{SS}}_{n,n^\prime\neq n}} |^2$.
In Fig.~\ref{fig:A_U_traj}(c) we compute the distance between $\overline{C^{\text{SS}}}$ obtained from the unitary evolution, and $C^{\text{ME}}$ obtained from the master equation (Eq.(2) of the main text), $\tilde{C}^{\text{SS}, \text{ME}}_{n, n^\prime} = \sum_{n,n^\prime} | C^{\text{SS}, \text{ME}}_{n, n^\prime} - \overline{C^{\text{SS}}_{n, n^\prime}}|^2$.
The deviation from the infinite-temperature state decreases as $N_{\text{init}}$ increases.
With $N_{\text{init}} = 256$, for example, $\overline{C^{\text{SS}}_{n, n}}/ \overline{C^{\text{SS}}_{n, n^\prime \neq n}} \approx 5 \times 10^2$ at $t=t_{\text{s}}$.

\end{document}